\documentclass[xcolor=pdftex,usenames,dvipsnames,table]{PoS}

\usepackage{slashed}
\usepackage{graphicx}
\usepackage{amssymb,amsmath}
\usepackage{pifont}
\usepackage{multirow,lscape}
\usepackage{array}
\usepackage[lofdepth,lotdepth]{subfig}
\usepackage{verbatim}
\usepackage{bm}
\usepackage{comment}

\graphicspath{{./figs/}}

\providecommand{\abs}[1]{\lvert#1\rvert}

\providecommand{\CL}{\nonumber\\}
\newcolumntype{C}[1]{>{\centering\arraybackslash}p{#1}}
\providecommand{\fwd}[1]{\textcolor{NavyBlue}{t_{#1}}}
\providecommand{\bck}[1]{\textcolor{BrickRed}{t_{-{#1}}}}

\title{Optimization of the Oktay-Kronfeld Action Conjugate Gradient
  Inverter}  

\ShortTitle{Optimization of the OK Action CG Inverter}

\author{%
  \speaker{Yong-Chull Jang}, Jon A. Bailey, Weonjong Lee$^1$ \\
  Lattice Gauge Theory Research Center, CTP, and FPRD, \\
  Department of Physics and Astronomy, 
  Seoul National University,
  Seoul, 151-747, South Korea\\
  E-mail: $^1$\email{wlee@snu.ac.kr}
}
\author{%
  Carleton DeTar$^{a,2}$, Mehmet B. Oktay$^{a,b}$\\
  $^a$Department of Physics and Astronomy, University of Utah,
  Salt Lake City, UT  84112, USA\\
  $^b$Department of Physics and Astronomy, University of Iowa,
  Iowa City, IA  52242, USA\\
  E-mail: $^2$\email{detar@physics.utah.edu}
}
\author{%
  Andreas S. Kronfeld\\
  Theoretical Physics Department,\\ Fermi National Accelerator Laboratory\thanks{%
Operated by Fermi Research Alliance, LLC, under Contract No.\ DE-AC02-07CH11359 with the United
States Department of Energy.},
  Batavia, IL  60510, USA\\
  E-mail: \email{ask@fnal.gov}
}
\author{SWME, MILC, and Fermilab Lattice Collaborations}
\abstract{%
Improving the Fermilab action to third order in heavy quark effective theory
yields the Oktay-Kronfeld action, a promising candidate for precise
calculations of the spectra of heavy quark systems and weak matrix elements
relevant to searches for new physics. We have optimized the bi-stabilized
conjugate gradient inverter in the SciDAC QOPQDP library and are developing a
GPU code.  The action is rewritten and the needed gauge-link combinations are
precalculated.  In tests with a MILC coarse lattice, this procedure accelerates
the inverter by a factor of four.  The remaining floating-point operations are
mostly simple matrix multiplications between gauge links and fermion vectors,
which we accelerate by more than an order of magnitude by using CUDA.  Further
gains could be achieved by using QUDA.
}

\FullConference{%
  31st International Symposium on Lattice Field Theory - LATTICE 2013\\
  July 29 - August 3, 2013\\ Mainz, Germany}

\begin{document}

\section{Introduction}
%

The quantity $\epsilon_K$ describes indirect CP violation in the
$K^0$-$\bar{K}^0$ system and enters tests of CKM unitarity and other searches for
new physics.  
The dominant sources of uncertainty in the Standard Model (SM) value of
$\abs{\epsilon_K}$ are, first, the theory uncertainty in $|V_{cb}|$, which stems from the form
factors calculated with lattice QCD, and, second, the uncertainty in the matrix element $\hat{B}_K$, also from 
lattice QCD.  
The values for $\hat{B}_K$ and $|V_{cb}|$ used in Ref.~\cite{Jang2012:PoS.LAT2012.269} were updated at this
conference~\cite{swmebk,JackVcb}.
With these results, the tension between the SM calculation
and the experimental measurement of $|\epsilon_{K}|$ remains in excess of $3\sigma$.
(With the inclusive value~\cite{GS} of $|V_{cb}|$, the tension vanishes.  The exclusive and inclusive values of $|V_{cb}|$ differ by $3\sigma$~\cite{JackVcb}.)

New lattice calculations of the form factors of the exclusive decays $\bar{B}\to D^{(*)}\ell\bar{\nu}$, 
which are used to determine~$|V_{cb}|$, are essential.
Heavy-quark discretization errors are the largest source of uncertainty at present, and the Oktay-Kronfeld 
(OK) action~\cite{cite:OK} has been designed to reduce them.
The OK action was developed by improving the Fermilab action~\cite{EKM} through third order in
HQET~\cite{cite:OK}.
With tree-level matching, the third-order improvement terms consist of four dimension-6 and two dimension-7
bilinears; no four-fermion operators arise.
The HQET analysis suggests that the charm-quark discretization errors of the OK action are 
comparable to those of other highly-improved actions, while bottom-quark discretization effects are 
smaller~\cite{cite:OK}.
In this report, we describe an optimized conjugate gradient (CG) inverter for the OK action.
For performance tests we use the tree-level, tadpole-improved action that gave encouraging, albeit
preliminary, results for the spectrum ~\cite{MBO:LAT2010}.

\section{Optimization}
\subsection{Dirac Operator}
For a Dirac operator $M$ and source vector $\xi$, the
system of equations
\begin{align}
  \sum_{y\beta b} M_{xy}^{\alpha\beta,ab} \psi_y^{\beta b} 
    &= \xi_x^{\alpha a}
  \label{eq:dslash}
\end{align}
must be solved to construct lattice correlators; $x$ and $y$ label the
lattice sites, $\alpha$ and $\beta$ are spin indices, and $a$ and
$b$ are color indices.
The solution vector $\psi$ can obtained by the CG method.
This algorithm iteratively updates the vector $\psi$ from an initial guess.
For each update the matrix multiplication of Eq.~(\ref{eq:dslash}) is
required.

We focus on optimizing this matrix multiplication by reducing
the number of floating-point operations.
We also consider how to exploit the size of local memory and node-to-node
communication speed to increase efficiency without sacrificing performance
gains from reducing the number of floating-point operations.

We first rewrite the OK action by collecting terms with products of gauge links
multiplying the same neighboring fermion field.  Suppressing spin and color
indices, 
\begin{align}
 \sum_{y} M_{xy} \psi_{y} = &\; W^{0}_{x} \psi_{x}
+\sum_{\mu} \Big( W^{+}_{\mu,x} \psi_{x+\hat{\mu}} +W^{-}_{\mu,x}
\psi_{x-\hat{\mu}} \Big) +\sum_{i} \Big( W^{++}_{ii,x} \psi_{x+2\hat{i}}
+W^{--}_{ii,x} \psi_{x-2\hat{i}} \Big) \CL
&+\sum_{j>i} \Big\{ W^{++}_{ij,x}
\psi_{x+\hat{i}+\hat{j}} +W^{--}_{ij,x} \psi_{x-\hat{i}-\hat{j}} +W^{+-}_{ij,x}
\psi_{x+\hat{i}-\hat{j}} +W^{+-}_{ji,x} \psi_{x-\hat{i}+\hat{j}} \Big\} \;,
\label{eq:minFlopForm} 
\end{align} 
where $\mu=1,2,3,4$, $\mu=4$ is the temporal direction, and $i$ runs over the
spatial indices.  For a given position $x$, each $W$ is a $12 \times 12$ matrix
in spin-color space.  $W$ consists of sixteen $3 \times 3$ color matrices.
They are sums of gauge-link products of up to 5 links and a constant.  Each
gauge-link product carries a factor of $\pm 1$ or $\pm i$ that depends on the
involved $\gamma_{\mu}$.

The $W$ matrices remain the same for all CG iterations.  We precalculate and
reuse them to accelerate the CG iteration.  In subsequent sections we call them
``precalculation matrices.''

\subsection{Precalculation}
Precalculation decreases the number of floating-point operations required for
the Dirac operation, but saving the entire set requires too much memory.  
The full set of precalculation matrices in Eq.~(\ref{eq:minFlopForm}) has $432$
color blocks.  The size of a gauge configuration is $4$ color blocks.  It turns
out that we do not need to hold everything in memory if we exploit the
conjugate relation between opposite direction pairs of precalculation
matrices.  After introducing explicit representations for $\gamma_{\mu}$,
we can see that some color blocks are the same or equal~zero.

Precalculation matrices multiplied to the off-diagonal($i \neq
j$) next-to-nearest neighbor fermion fields satisfy the following
relations.
\begin{equation}
  W^{--}_{ij,x} 
    = -W^{++ \dagger}_{ij,x-\hat{i}-\hat{j}} \;,\quad
  W^{+-}_{ji,x} 
    = -W^{+- \dagger}_{ij,x-\hat{i}+\hat{j}} \;.
  \label{eq:conjugate}
\end{equation}
%
%
The operation of Hermitian conjugation is applied in both spin and color
spaces.  Although the necessary relations are more complicated, the diagonal
precalculation matrices $W^{-}_{\mu,x}, W^{--}_{ii,x}$ can be obtained from
their positive direction counterparts $W^{+}_{\mu,x-\hat{\mu}},
W^{++}_{ii,x-2\hat{i}}$.  Hermitian conjugation, a sign change, and color-block
reordering are required, depending on the representation chosen for the Dirac
matrices $\gamma_{\mu}$.  The relations are given explicitly in
Eq.~(\ref{eq:opp_diag_pre_mat}).

Hence, excepting $W^{0}_{x}$, the memory requirement for the precalculation
matrices can be reduced by a factor of two.  In the end, we can cut the
required memory down to $50$ color blocks, excluding the identical and
vanishing color blocks.\footnote{The mass term in $W^{0}_{x}$ is separately
treated in practice.  It saves memory by $1$ more color block, instead of
increasing the number of floating-point operations.} As a by-product, the
unnecessary floating-point operations required for constructing the
precalculation matrices and for multiplying the fermion fields by null color blocks are removed.  

\begin{figure}[b]
\centering
\includegraphics[width=0.7\textwidth]{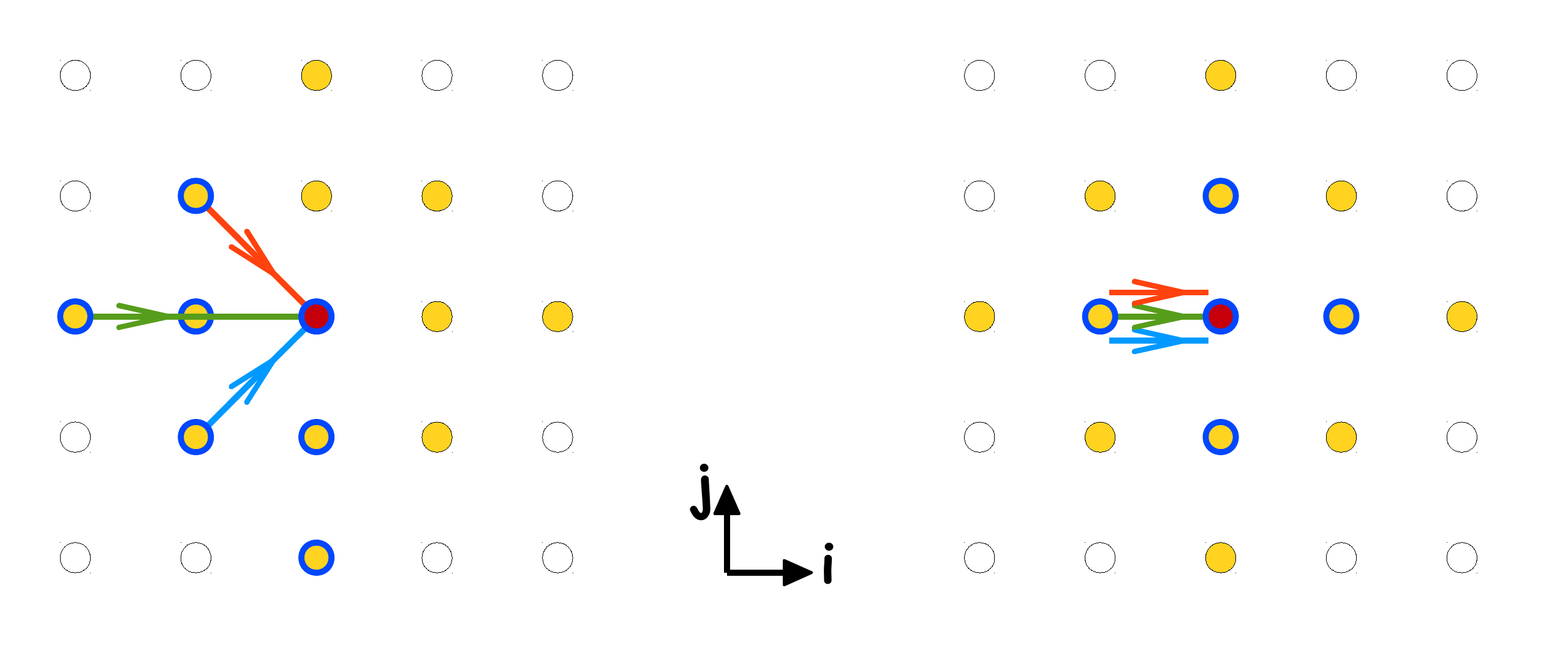}
\caption{To update the fermion field on the site $x$ (red), one needs in
addition the fermion fields defined on the neighboring sites (yellow).  The
precalculation matrices defined on the blue-circled sites are also needed.  By
calculating shifted precalculation matrices, the field access pattern is
simplified, saving floating-point operations.}
\label{fig:W_pattern}
\end{figure}

Though beneficial in terms of reducing floating-point operations, exploiting
the conjugation relations introduces a complicated field access pattern because
the operation of Hermitian conjugation is applied to the precalculation matrix
shifted in the opposite direction.  This pattern can be seen in
Eqs.~(\ref{eq:conjugate}) and (\ref{eq:opp_diag_pre_mat}).  To update the
fermion field on the site $x$, the fermion fields on the neighboring sites need
to be collected to the site $x$.  Then these and the on-site fermion field are
multiplied by the pair of precalculation matrices.  (In the temporal direction,
only the nearest neighbors are involved.  In the spatial directions, all the
nearest and next-to-nearest neighbors participate.)  However, using the
conjugation relations requires collecting not only the fermion fields, but also
the precalculation matrices, which reduces off-node performance.  Copying the
precalculation matrices can be avoided by simplifying the access pattern.  We
distribute precalculation matrices multiplied by a next-to-nearest neighbor
fermion field over the nearest neighbors.  This simplification is depicted in
Fig.~\ref{fig:W_pattern}.

The shifted precalculation matrices can be identified by rewriting 
Eq.~(\ref{eq:minFlopForm}).  We have
\begin{align}
  \sum_{y} M_{xy} \psi_{y} =
    &\; W^{0}_{x} \psi_{x}
     +\sum_{\mu} \Big( W^{+}_{\mu,x} \psi_{x+\hat{\mu}} 
           + \bck{\mu} W^{-}_{\mu,x+\hat{\mu}} \psi_{x} \Big)
     +\sum_{i} \Big( \fwd{i} W^{++}_{ii,x-\hat{i}} \psi_{x+\hat{i}} 
                   + \bck{i} W^{--}_{ii,x+\hat{i}} \psi_{x-\hat{i}} \Big) \CL
    &+\sum_{j>i} \Big\{
        \fwd{j} W^{++}_{ij,x-\hat{j}} \psi_{x+\hat{i}}
       -\bck{i} W^{++ \dagger}_{ij,x-\hat{j}} \psi_{x-\hat{j}}
       +\bck{j} W^{+-}_{ij,x+\hat{j}} \psi_{x+\hat{i}}
       -\bck{i} W^{+- \dagger}_{ij,x+\hat{j}} \psi_{x+\hat{j}} \Big\} \;,
  \label{eq:FinalForm}
\end{align}
where $t_{\pm \mu}$ are translation operators that shift the function (field)
$f_x$ by one lattice spacing in each direction.
\begin{align}
  t_{\pm \mu} f_x = f_{x \pm \hat{\mu}} \;.
\end{align}
The set of precalculation matrices saved for site $x$ consists of $W^{0}_{x},
W^{+}_{\mu,x}, W^{++}_{ii,x-\hat{i}}, W^{++}_{ij,x-\hat{j}}$ and
$W^{+-}_{ij,x+\hat{j}}$, $(i<j)$.  Using the $\gamma_{\mu}$ representation
in~Ref.~\cite{EKM} and the notation of Ref.~\cite{MBO:LAT2010}, the explicit
form of the precalculation matrices is
\begin{gather}
  W^{0}_{x}
    =  \frac{u_0}{2\kappa}
      +\begin{pmatrix}
         D^0_x & S^0_x \\
        -S^0_x & D^0_x
       \end{pmatrix}\,,\quad
  D^{0}_{x}
    =-\frac{c_B \zeta + 16 c_5}{2 u_0^3} \bar{B}^{D}_x \,,\quad
  S^{0}_{x}
    = -\frac{c_E \zeta}{2 u_0^3} \bar{E}^{S}_x\,, 
    \label{eq:shifted_W_first}\\
  \bar{E}^{S}_x
    = \sum_{i=1}^{3} \sigma_i E_{i,x}\,,\quad
  \bar{B}^{D}_x
    = i \sum_{i=1}^{3} \sigma_i B_{i,x}\,,\quad
  W^{+}_{4,x}
    = \begin{pmatrix}
         0 & S^{+}_{4,x}  \\
         S^{+}_{4,x} & -U_{4,x}
       \end{pmatrix} \,,\quad
  S^{+}_{4,x}
    =  \frac{c_{EE}}{2 u_0^4} 
       ( U_{4,x} \bar{E}^{S}_{x+4} - \bar{E}^{S}_x U_{4,x} )\,, \\
  W^{+}_{i,x}
    = \begin{pmatrix}
         D^{+}_{i,x} & S^{+}_{i,x}  \\
         S^{+}_{i,x} & D^{+}_{i,x}
       \end{pmatrix} \,,\quad
  S^{+}_{i,x}
    = \frac{1}{2} ( \zeta - 2 c_1 - 12 c_2 ) \sigma_i U_{i,x}
      +\frac{c_3}{2 u_0^4}
       (  U_{i,x} \sigma_i \bar{B}^{D}_{x+i}
        - \sigma_i \bar{B}^{D}_{x} U_{i,x}
        + 2i B_{i,x} U_{i,x} ) \,,\CL
  D^{+}_{i,x}
    = -\frac{1}{2}( r_s \zeta + 8 c_4 ) U_{i,x}
      +i\frac{c_5}{4} \left(u_0^{-2} - u_0^{-4}\right)
       \sum_{j,k=1}^{3} \epsilon_{ijk} \sigma_{j}
       ( U_{k,x} U_{i,x+k} U^{\dagger}_{k,x+i}
        -U^{\dagger}_{k,x-k} U_{i,x-k} U_{k,x-k+i} ) \CL
      +\frac{c_5}{u_0^4}
       \left[  U_{i,x} \bar{B}^{D}_{x+i} + \bar{B}^{D}_x U_{i,x}
        -i \sigma_i (U_{i,x} B_{i,x+i} + B_{i,x} U_{i,x}) \right]\,,\\
  W^{++}_{ii,x}
    =  \begin{pmatrix}
         D^{++}_{ii,x} & S^{++}_{ii,x} \\
         S^{++}_{ii,x} & D^{++}_{ii,x}
       \end{pmatrix} \,,\quad
  D^{++}_{ii,x}
    = \frac{c_4}{u_0} U_{i,x} U_{i,x+i} \,,\quad
  S^{++}_{ii,x}
    = \frac{c_1 + 2 c_2}{2 u_0} \sigma_i U_{i,x} U_{i,x+i}\,,\\
  W^{-}_{4,x}
    =  \begin{pmatrix}
         -U^{\dagger}_{4,x-4} & -S^{+ \dagger}_{4,x-4} \\
         -S^{+ \dagger}_{4,x-4} & 0
       \end{pmatrix} \,,\quad
  W^{-}_{i,x}
    =  \begin{pmatrix}
         D^{+ \dagger}_{i,x-i} & -S^{+ \dagger}_{i,x-i} \\
         -S^{+ \dagger}_{i,x-i} & D^{+ \dagger}_{i,x-i}
       \end{pmatrix} \,,\quad
  W^{--}_{ii,x}
    =  \begin{pmatrix}
         D^{++ \dagger}_{ii,x-2i} & -S^{++ \dagger}_{ii,x-2i} \\
        -S^{++ \dagger}_{ii,x-2i} &  D^{++ \dagger}_{ii,x-2i}
       \end{pmatrix}\,,
  \label{eq:opp_diag_pre_mat}\\
  W^{++}_{ij,x}
   =  \begin{pmatrix}
         0 & S^{++}_{ij,x} \\
         S^{++}_{ij,x} & 0
       \end{pmatrix} \,,\quad
  S^{++}_{ij,x}
    = \frac{c_2}{2 u_0} ( \sigma_i + \sigma_j )
       ( U_{i,x} U_{j,x+i} + U_{j,x} U_{i,x+j} ) \,,\;
  (i \neq j)\,,\\
  W^{+-}_{ij,x}
    =  \begin{pmatrix}
         0 & S^{+-}_{ij,x} \\
         S^{+-}_{ij,x} & 0
       \end{pmatrix} \,,\quad
  S^{+-}_{ij,x}
    = \frac{c_2}{2 u_0} ( \sigma_i - \sigma_j )
      ( U_{i,x} U^{\dagger}_{j,x-j+i} + U^{\dagger}_{j,x-j} U_{i,x-j} )
      \,,\;
  (i \neq j)\,.
  \label{eq:shifted_W_last}
\end{gather}

\begin{figure}[b]
\centering
\includegraphics[width=0.7\textwidth]{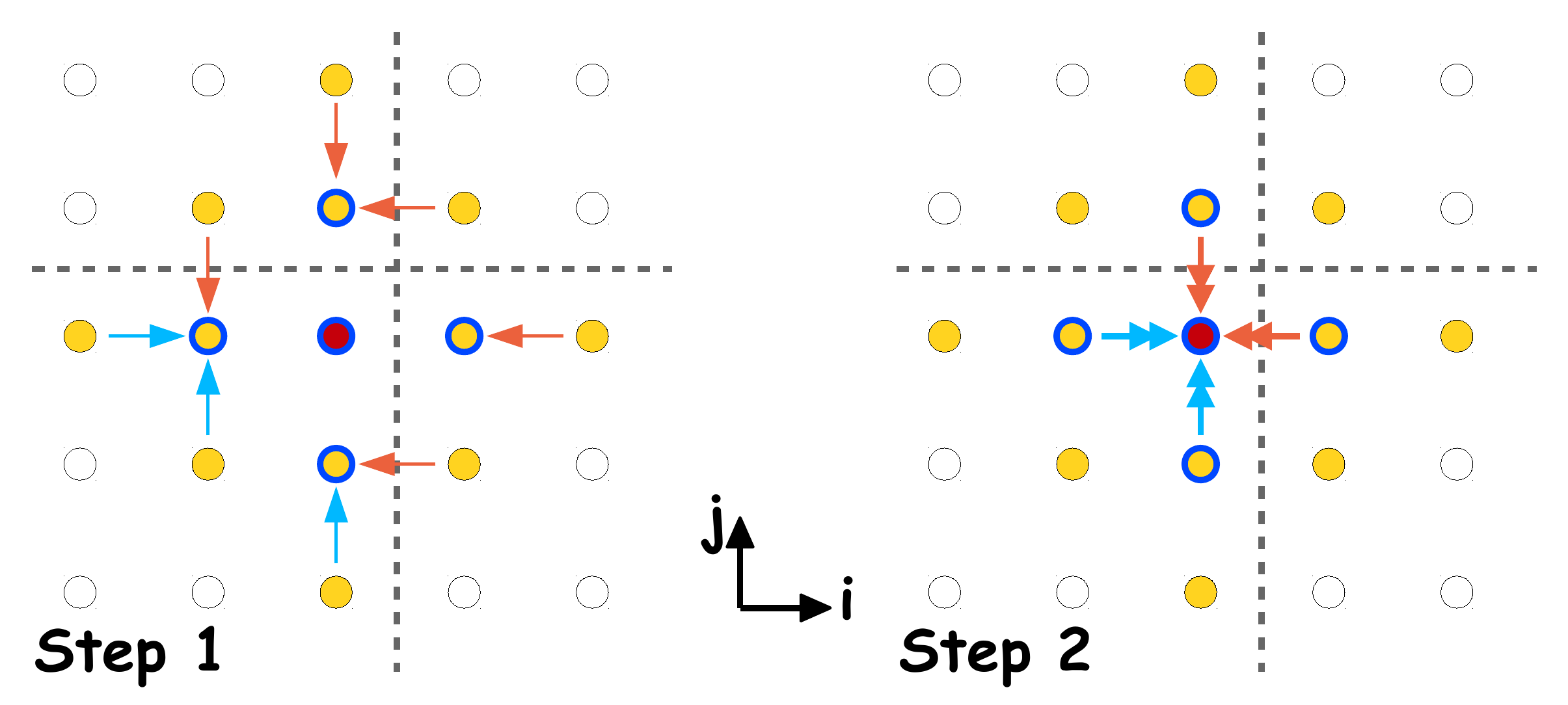}
\caption{The dashed line is the computing node boundary.
To update the fermion field (red),
in the first step, only the nearest neighbor fermion fields are gathered and
multiplied with the precalculation matrices (blue circles).
In the second step, the resulting products are gathered from the nearest
neighbor sites and added together.}
\label{fig:comm_pattern}
\end{figure}

The matrix multiplications in Eq.~(\ref{eq:FinalForm}) are isolated from shift
operations occurring before and after the multiplications.  Pre-multiplication
shifts gather nearest neighbor fermion fields.  Post-multiplication shifts
distribute the multiplication results to the nearest neighbors, followed by the
fermion vector sum.  Through these steps, shown in Fig.~\ref{fig:comm_pattern},
each next-to-nearest neighbor fermion field contribution is propagated to the
proper destination site.  In a parallel computing environment, such as MPI,
off-node communications are necessary only for the outermost boundary surfaces.

\subsection{Implementation}
In the USQCD library~\cite{usqcd}, a test version of the OK action CG inverter
was included as part of QOPQDP.  The inverter consists of a general purpose
QOPQDP inverter that uses the bi-stabilized CG algorithm and a specific
implementation of the OK action Dirac operator.  The MILC library serves as our
testing environment for the OK action CG inverters.  We perform a mixed precision inversion by calling the QOPQDP inverter or Dirac operation module in single or double precision, as appropriate.

For CPU clusters, the optimized OK Dirac operator of Eq.(\ref{eq:FinalForm}) is
implemented as part of QOPQDP.  For GPU clusters, only part of the matrix
multiplication in the optimized Dirac operator is replaced with CUDA function
calls.  To write working GPU code, we do not need to alter the parts of the
optimized CPU code responsible for communication and precalculation.  As
reflected in the performance results, this GPU module can be further optimized.

\section{Performance}
To measure CG performance, we use a MILC coarse ($a\approx0.12\ \mathrm{fm}$)
lattice with dimensions $20^{3} \times 64$.  The lattice is divided to fit 4
nodes of the SNU cluster DAVID1.  Each node consists of one core of an Intel
i7-920 CPU together with an NVIDIA GTX480 GPU.  Each node communicates with a
single-rail QLogic InfiniBand network.

Precalculation reduces overall CG time by a factor of $3.9$ times.  This gain
is increased to $13.1$ when the precalculation matrix multiplication is
performed with CUDA.  (See Table~\ref{table:CGtime}.)
Table~\ref{table:mm_timing} and \ref{table:c_timing} show the timing details
and CG performance in GFLOPS.  Counting only the matrix multiplication, the
maximum performance is $58.7$ GFLOPS.  Including memory copy time between host
and GPU global memory for the precalculation matrix and the fermion field, the
performance is decreased to $18.2$ GFLOPS.  

Because the CUDA module is called from the QOPQDP side, another
overhead of QOPQDP preparation time arises.  To pass the QDP data
types to the external CUDA function, they should be exposed to the C
intrinsic pointer variables.  After arithmetic on the GPU, they must
be recast to the previous QDP data types.  With the current
implementation, the total CUDA overhead time, which consists of memory
copy time (= CUDA Memory Copy, $W$ + CUDA Memory Copy, $\psi$ in Table
\ref{table:mm_timing}) and QOPQDP preparation time, exceeds the matrix
multiplication time by a factor of 3.7 (5.1) for a single Dirac
operation of single (double) precision.
\begin{table}[tb]
  \footnotesize 
  \tabcolsep 3pt
  \centering
  \renewcommand{\arraystretch}{1.2}
  \subfloat[][CG Time]{
  \resizebox{.25\textwidth}{!}{
    \begin{tabular}{r|r|r}
    \hline\hline
    \multicolumn{1}{c}{Naive}&
    \multicolumn{1}{|c}{Precalc.}&
    \multicolumn{1}{|c}{CUDA}\\[0.2ex]
    \hline\hline
    11814.8& 3048.8& 898.7\\ \hline\hline
    \end{tabular}
  }
  \label{table:CGtime}
  }
  \subfloat[][Matrix Multiplication]{
  \resizebox{.43\textwidth}{!}{
    \begin{tabular}{l|r|r}
    \hline\hline
    \multicolumn{1}{c}{}&
    \multicolumn{1}{|c}{Precalc.}&
    \multicolumn{1}{|c}{CUDA}\\[0.2ex]
    \hline\hline
    Matrix Multiplication& 962[1206]& 32[107]\\ 
    \hline
    CUDA Memory Copy, $W$& & 4[262]\\ \hline
    CUDA Memory Copy, $\psi$& & 60[121]\\ \hline
    QOPQDP Preparation& & 54[163]\\ \hline\hline
    \end{tabular}
  }
  \label{table:mm_timing}
  }
  \subfloat[][Common Module]{
  \resizebox{.28\textwidth}{!}{
    \begin{tabular}{l|r}
    \hline\hline
    Communication& 2[11]\\ \hline
    Gamma Basis Change& 16[32]\\ \hline
    Spin Decomposition& 23[45]\\ \hline
    Vector Addition& 69[66]\\ \hline\hline
    \end{tabular}
  }
  \label{table:c_timing}
  }
\caption{ CG performance: (a) The CG time is measured in
  seconds. ``Naive'' means the original CG inverter without any
  improvement. ``Precalc'' means the CG inverter with precalculation
  of $W$ matrices. ``CUDA'' means the CG inverter with precalculation
  and with the Dirac opertor programmed in CUDA. (b) The values are
  measured in the unit of milliseconds. Each value represents the time
  elapsed per single CG iteration. Values in the bracket $[\cdots]$
  correspond to the double precision calculation. (c) The same
  notation as in (b).}
\end{table}

The GTX480 has $1.5$ GB of global memory, which is not large enough to
hold all the necessary precalculation matrices at once.  At best we
can allocate GPU global memory space for the full set of single
precision precalculation matrices.  The double precision update is
divided into two parts so that the double precision precalculation
matrices can be held by the allocated GPU global memory space.  For
each double precision update, the sets of precalculation matrices are
copied in succession from the host memory.  Single precision
precalculation matrices are copied at the beginning of the iterations
and used again in each iteration.  When the precision is changed, the
precalculation matrices for the other precision are wiped from the GPU
global memory.  This precalculation matrix copy overhead can be
overcome by using a GPU (such as GTX Titan) with global memory
sufficient to store the single and double precision precalculation
matrices.

The remaining overheads are the fermion field copy time (CUDA Memory
Copy, $\psi$ in Table \ref{table:mm_timing}) and QOPQDP preparation
time.  Together these occupy $96.6\%$ ($52.0\%$) of the total CUDA
overhead time in single (double) precision, so we expect more
optimization of the GPU inverter can be achieved without new hardware.

\section{Future Work}
By developing the OK action inverter with QUDA, the QOPQDP preparation time can
be removed.  Reducing the CPU-GPU communication time requires reducing
overheads from copying the fermion fields and the precalculation matrices; the
latter could be addressed with hardware with more global memory.  Finally,
$M^{\dag}M$ preconditioning, even-odd preconditioning, and spin projection are
commonly used to optimize inversions of the Dirac operator with the CG
algorithm.  Even-odd preconditioning appears very difficult for the OK action;
we have not yet investigated how to implement the remaining two techniques.

\section*{Acknowledgments}

This work was supported in part by the U.S.\ Department of Energy under grant No.\ DE-FC0212ER-41879 (C.D.)
and the U.S.\ National Science Foundation under grant PHY10-67881 (C.D.).
The research of W.~Lee is supported by the Creative Research Initiatives Program (2013-003454) of the NRF
grant funded by the Korean government (MSIP).
W.~Lee would like to acknowledge the support from KISTI supercomputing center through the strategic support
program for the supercomputing application research [No.~KSC-2012-G3-08].
Computations were carried out on the DAVID GPU clusters at Seoul National University.
The research of J.A.B.
is supported by the Basic Science Research Program (2013009149) of the National Research Foundation of Korea
(NRF) funded by the Ministry of Education.
%

\end{document}